\documentclass{aa}
\usepackage{txfonts,psfig}

\begin{document}
\title{Blank Field Sources in the ROSAT HRI Brera Multiscale Wavelet catalog}
\author{
M. Chieregato\inst{1,2}
\and
S. Campana\inst{3}
\and 
A. Treves\inst{1}
\and
A. Moretti\inst{3}
\and 
R.P. Mignani\inst{4}
\and 
G. Tagliaferri\inst{3}
}

\authorrunning{Chieregato et al.}
\titlerunning{BFS in the Rosat BMW-HRI catalog}
\offprints{Matteo Chieregato}

\institute{Universit\`a dell'Insubria,
               Via Valleggio 11, I-22100 Como, Italy \\
              \email{matteo.chieregato@uninsubria.it}          
             \and
             Istitut f\"ur Theoretische Physik der Universit\"at Z\"urich,
Winterthurerstrasse 190, CH-8057, Z\"urich, Switzerland
             \and
             Osservatorio Astronomico di Brera, Via Bianchi 46, I-23807 Merate
(LC), Italy
          \and
European Southern Observatory, Karl-Schwartzschild Strasse 2, D-85740 Garching, Germany
              }

   \date{Received; accepted}


\abstract{
The search for Blank Field Sources (BFS), i.e. X-ray sources
without optical counterparts, paves the way to the identification of unusual
objects in the X-ray sky. Here we present four BFS detected in the Brera
Multiscale Wavelet catalog of ROSAT HRI observations. This sample has been
selected on the basis of source brightness, distance from 
possible counterparts at other wavelengths, point-like shape and good
estimate of the X-ray flux ($\rm f_X$). 
The observed $\rm f_X$ and the limiting
magnitude of the optical catalogs fix a lower limit for our BFS on ${\rm
f_X/f_{opt}}$ $\sim$ 40. This value puts them well beyond 90$\%$ threshold for
usual source classes once HRI energy band and proper spectral shape
are taken into account, leaving room for speculation on their
nature. Three BFS show also evidence of a transient behaviour.

}

\maketitle

\section{Introduction}

It is well known that one can discriminate among classes of X-ray sources
evaluating the ratio between their fluxes in X-ray and optical bands
(e.g. Maccacaro et al. \cite{macca1988}; Motch et al. \cite{motch1998};
Zickgraf et al. \cite{zick2003}), especially if some spectral information is
supplemented. 
Blank Field Sources (BFS), i.e. X-ray sources with no optical counterparts,
stand somehow apart from the bulk of X-ray zoology in that they represent
rarer, less studied objects (e.g. Cagnoni et al. \cite{ila2002}). 
In particular, known X-ray sources concentrate at not too high values of ${\rm
f_X/f_{opt}}$, e.g., stars have ${\rm f_X/f_{opt}} \lesssim $1, galaxies and AGN reach ${\rm
f_X/f_{opt}}$ of a few tens. At larger values of ${\rm f_X/f_{opt}}$, peculiar
populations start to show up: some examples are isolated neutron stars,
type 2 quasars, extreme BL Lacs or high-redshift clusters of galaxies. 

The search of BFS aims at objects with high ${\rm f_X/f_{opt}}$: the observed
X-ray flux and the upper limit from the absence of an optical detection
establish a lower limit on the ${\rm f_X/f_{opt}}$. A careful selection of
bright BFS is therefore a powerful tool for finding population leftovers from
usual source classes. 
The status of BFS is somehow a transitory condition that expresses our lack of
knowledge on the real nature of these sources. New X-ray observations can
provide temporal and spectral information on BFS, while deeper optical/near IR
pointings can either reveal the counterpart or establish a much stronger lower
limit on the ${\rm f_X/f_{opt}}$. 
With the aid of these information, BFS can be identified and classified
properly. 

In this paper we present a sample of luminous BFS from the ROSAT HRI Brera
Multiscale Wavelet catalog (BMW-HRI: Panzera et al. \cite{panz2003}; Campana
et al. \cite{camp1999}; Lazzati et al. \cite{lazz1999}), with ${\rm
f_X/f_{opt}}\gtrsim 40$ \footnote{Actually this value is ${\rm f_{0.5-2
keV}/f_{BJ}}$, see Section 2.}, along with archival X-ray spectral data from
other observations (when available). This is the first study to exploit all
ROSAT HRI data for BFS search (see Musso et al. \cite{musso1998} for the first
search on a much smaller dataset). A key factor in the identification process
is the uncertainty of the X-ray position (see Cagnoni et al. \cite{ila2002};
Rutledge et al. \cite{rut2003}); on this respect, the use of HRI rather than
ROSAT PSPC images for searching BFS is much superior.

The paper is organized as follows: in Section 2 we briefly describe the
BMW-HRI catalog and its cross-identification program with optical, infrared
and radio catalogs, in Section 3 we introduce the method of sample selection,
while the final sample of BFS is presented in Section 4. 
In Section 5 we discuss the possible nature of BFS and future
search plans are mentioned in Section 6.

\section{The BMW-HRI catalog}

\subsection{Catalog description}

The BMW-HRI catalog (Panzera et al. \cite{panz2003}) is derived from
an analysis of the ROSAT HRI data set with a source detection algorithm based
on the wavelet transform (Lazzati et al. \cite{lazz1999}; Campana et
al. \cite{camp1999}).  

The up-to-date version of the catalog contains 28998 entries, down to a count
rate of $\sim 10^{-4}$ cts s$^{-1}$. The total sky coverage is 732 deg$^2$
($\sim$ 1.8$\%$ of the sky). 
For each entry name, position, count rate, extension and relative errors are
provided, along with derived parameters like flux and Galactic column density
and ancillary information about the pointing. Furthermore, results of
cross-correlations with GSC2, 2MASS, IRAS, and FIRST catalogs are reported
(see Section 2.2 and Table \ref{crossid}). In the following, some of the basic
characteristics of the catalog are discussed in order to permit an
understanding of the BFS selection method. Further details can be found in
Panzera et al. (\cite{panz2003}).  

\textbf{Extension.} The wavelet transform is particularly powerful in dealing
with extended sources. In practice, each detected source is characterized by
its extension, alongside with the more usual parameters (count rate and
position). Extended sources, especially when observed at large off-axis
angles, are often broken in various point-like sources by other
detection algorithms. This risk is minimized by the wavelet transform.

\textbf{Detection threshold and spurious sources.} 
Each BMW-HRI source is characterized also by a detection probability value
(peak significance). This value is the probability, in units of Gaussian
$\sigma$, that the source is not spurious (i.e. a background fluctuation). 
This probability is the result of numerical simulations of random fields
(without sources), repeated for different wavelet transform scales; it takes
into account the number of background peaks mismatched as sources by the
detection algorithm, for a given signal to noise ratio threshold in wavelet
space. The mean number of spurious sources expected over an image is kept
constant (0.4 spurious detections per field), varying the detection threshold.
That is, the detection threshold depends on the mean background counts
value on the image and on the scale at which the search is performed (see
Lazzati et al. \cite{lazz1999}). The mean threshold value in the
overall catalog is $\sim 4.2\sigma$. 
  
\textbf{Errors.} In principle, the error for each parameter found with the
wavelet transform algorithm can be estimated from the covariance matrix. In
practice, such errors are reliable only when the number of source and
background counts are sufficiently high ($\gtrsim 2 \times 10^{-2}$ counts per
pixel). Otherwise, the distribution of the wavelet transform coefficients
becomes Poissonian rather than Gaussian (in the lowest scales). 
If this is the case, a better estimate of the errors can be given by means of
basic statistics (such as the error on the number of counts $N$ is $\sqrt{N}$,
for more details see Lazzati et al. \cite{lazz1999}). 
The errors reported here and in the catalog are the maximum between the
covariance matrix and the statistical estimates by Lazzati et al. (\cite{lazz1999}).

\textbf{Boresight correction uncertainty.}
There is an additional source of errors on the absolute position determination
of HRI sources, the uncertainty on the boresight correction. 
If the alignment between the node and the telescope optical axis is perfect,
on-axis images are exactly in the center of the instrument. However, in the
real situation there is always a finite misalignment that needs to be
corrected. Unfortunately, uncertainties in the aspect solution (that describes
the orientation of the telescope as a function of time) and errors in the
alignment between the star trackers and the ROSAT X-ray Telescope 
introduce uncertainties in the boresight correction, of variable size for each
observation. In practice, the systematic offset between accurately known
optical positions and X-ray positions can be used to evaluate the extent of
boresight correction uncertainties. For the ROSAT HRI, the offset can be as
large as $10''$, even if in most cases it will be much less (David et
al. \cite{david1998}). The $10''$ value is conservatively assumed as a
fiducial value for the boresight offset. Usually, the statistical errors on
the position from the detection algorithm, calculated as in the previous
section, are much less than $10''$ and therefore can be neglected in a first
approximation.  

\textbf{Multiple detections.} As each pointing is treated separately, not all
of the catalog entries correspond to independent sources. An estimate of the
overall number of independent sources can be given compressing the catalog
in a $10''$ radius (again, the fiducial error value indicated by the
boresight uncertainty). This procedure selects only a source for each $10''$
cone radius, on the basis of an autocorrelation of the position. With this
procedure, 20433 sources are left. Obviously, the compression brings to the
loss of sources truly close to each other. 

\subsection{Multiwavelength catalog cross correlations}

In the following, we will largely use the term \textit{off-band} to indicate a
wavelength passband different from X-rays. The limiting flux of a given
catalog in a given band will be generically denominated $\rm f_{offband}$.
For each of the BMW-HRI entries, cross correlations with GSC2 (McLean et
al. \cite{mclean2005}), 2MASS Second Data Release (Kleinmann et
al. \cite{klein1994}), IRAS Point Source Catalog (Beichman et
al. \cite{beich1988}) and FIRST Survey Catalog (White et al. \cite{white1997})
were performed by Panzera et al. (\cite{panz2003}). Off-band catalog
properties are summarized in Table \ref{crossid}.
The adopted radius for the cross-correlation is $10''$ (see Section 2.1), 
assumed as positional X-ray uncertainty. In fact, the positional uncertainties
for GSC2, 2MASS SDR and FIRST catalogs are $<0.5''$ ($3\sigma$,
McLean et al. \cite{mclean2005}), $<0.5''$ ($1\sigma$, see the 2MASS
documentation for an extensive discussion), $<0.5''$ ($90\%$, McMahon et
al. \cite{mcmahon2001}) respectively, so that they can be safely neglected for
cross-identification purposes.
An exception is the case of IRAS PSC, for which the positional accuracy varies
with source size, brightness and spectral shape and it is different in
different directions, but it is usually better than $20''$ (see Beichman et
al. \cite{beich1988}), so this last value has been used as cross-correlation
radius with this catalog. 
Note that: a) when two or more entries in the correlating catalog are found, 
only the brightest source parameters are reported (but in any case the number
of found sources is given), b) the sky coverage of the off-band catalogs is
usually not complete (see again Table \ref{crossid}). 
In particular, the preliminary, unpublished version of the GSC2 (GSC2.3)
catalog used in Panzera et al. (\cite{panz2003}) lacks coverage of the
zones where a bright source caused an overexposure of the Schmidt plates. For
the 2MASS and FIRST, more complete catalogs are now available (2MASS All Sky
Data Release, FIRST Survey Catalog 03Apr11 Version). Our final list has been
checked with them.

\begin{table}[htb]
\caption{{BMW-HRI cross-identifications.}\label{crossid}}
\vspace{0.1 cm}
\footnotesize
\begin{tabular}[c]{|l|l|l|l|l|}
\hline
Catalog       & Survey & Band & Coverage & Depth\\
\hline
GSC2          & POSSII & BJ & Dec: $>0$  & 22.5 \\
(prel.)       & ``     & R  & Dec: $>0$  & 20.8 \\
              & ``     & I  & Dec: $>0$$^*$ & 19.5 \\
              & SERC   & BJ & Dec: $<0$ & 23 \\
              & `` +AAO& R  & Dec: $<0$ & 22 \\
              & ``   & I  & Dec: $<0$$^*$ & 19.5 \\
\hline
2MASS SDR     & 2MASS  & J  & 50$\%$ sky & 15.8 \\
 (2000)       &        & H  & 50$\%$ sky & 15.1 \\
              &        & K  & 50$\%$ sky & 14.3 \\
\hline
IRAS PSC      & IRAS   & 12 $\mu$m & 98$\%$ sky & 0.4 Jy\\
(1989)        &        & 25 $\mu$m & 98$\%$ sky & 0.5 Jy\\
              &        & 60 $\mu$m & 98$\%$ sky & 0.6 Jy\\
              &        &100 $\mu$m & 98$\%$ sky & 1.0 Jy\\
\hline
FIRST SC      & FIRST  & 1.4 GHz &  20$\%$ sky & $\sim$ 1 mJy\\
(2001)        &        &         &             & \\
\hline
\end{tabular}
\vspace{0.2 cm}

$^*$ work in progress.
\par\noindent
\end{table}

\section{Sample selection}

\subsection{ {Catalog mask}}

The first sample selection was made on the basis of the following criteria:
\begin{enumerate}
\renewcommand{\theenumi}{\alph{enumi}}
\item blank field sources, i.e. without any cross identification in the other 
databases (see Section 2.2 and Table \ref{crossid}): 6061 catalog entries;
\item point-like sources (significance of the extension=0): 4955 catalog entries;
\item bright sources, i.e. with $f_{X} \ge 2.7 \times 10^{-13}$ erg s$^{-1}$
cm$^{-2}$: 275 catalog entries. 
\end{enumerate}
The reported numbers are catalog entries, i.e. no compression was applied to
eliminate multiple detections (see Section 3.2 below). The $\rm f_{X}$  we used is
computed from the observed source counts, in the 0.5-2 keV range, considering
only channels 2-9 (e.g. David et al. \cite{david1998}).  
The assumed spectral shape is a power law with photon index $\Gamma$=2
(i.e. the same of the Crab pulsar and nebula) and column density
$5\times10^{19}$ atoms cm$^{-2}$ (parameter $\#$ 48 in the BMW-HRI catalog).  
The chosen model for the count rate-flux conversion uses constant negligible
column density, instead of the integrated Galactic one in the source
direction. While the latter is commonly adopted in extragalactic surveys, it
would introduce a larger error on low-latitude galactic BFS, which are
expected to be low-luminosity, nearby objects (see Section 5). Furthermore,
the chosen $\rm f_{X}$ derived with this choice is lower than the one with
Galactic N$\rm _H$, so the obtained $\rm f_{X}/f_{offband}$ limit is lower. 
We chose to select sources on the basis of $\rm f_X$ and not of,
for example, $\rm f_X/f_{BJ}$. This brings naturally to different
$\rm f_X/f_{offband}$ ratios for sources in different parts of the sky,
due to the variable depths of the surveys available to us.
However, in such a way the selection is based on a directly observed quantity,
the number of counts, not on the basis of a non-detection, and it provides a
more homogeneous sample for future observations.
We carefully evaluated the $\rm f_X/f_{offband}$ keeping into account the
different depths and photometric band used in the different catalogs.
The obtained lower limit for the $\rm f_X/f_{opt}$ from the GSC2 catalog are
$\rm f_X/f_{BJ}>37$ (Northern sky), $\rm f_X/f_{BJ}>59$ (Southern sky),  
$\rm f_X/f_{RF}>43$ (Northern sky), $\rm f_X/f_{RF}>129$ (Southern sky). A
detailed description of method and used values can be found in Chieregato
(\cite{chie2005}).   

In conclusion, at odds with the majority of similar studies, our sources are
blank at various wavelengths (see again Table \ref{crossid}). We note also
that, although we selected point-like sources, we cannot exclude that the real
nature of some source is extended (i.e. only the brightest peak has been
detected with HRI).

\subsection{Multiple detections removal}

The removal of multiple detections from the sample could be done, in principle,
using automatic compression criteria (position based), like in Section 2.1.
However, given the limited entity of our sample of BFS candidates (275
elements), we preferred to remove multiple detections by hand to avoid, when
possible, the loss of sources truly close to each other.

We used positional coincidence as the main criterion for identifying multiple
detections. Another useful piece of information was the flux level. Unfortunately,
positional coincidence can be weakened by the boresight uncertainty, while
obviously the flux criterion is not fulfilled for highly variable sources. 
In practice multiple detections were reasonably identified in most of the cases. 
After multiple detections removal, we are left with 226 catalog entries.

\subsection{Total counts cut}

We made a further screening on the number of total counts, rejecting sources
with less than 25 photons (on the original detection, i.e. before
applying vignetting and Point Spread Function loss corrections). This has
been done in order to strengthen the derived source parameters (obviously, due
to the statistics increase). After the total counts cut, we are left with 201
sources. 

\subsection{Are the BFS sources real?}

We re-evaluated the number of spurious sources expected in this 201 subsample,
again on the basis of simulations of random fields with the above masks. 
In particular, the flux cut is very effective in the removal of
spurious sources.
In fact, the expected number of spurious source is $\sim 2$ in the 201
subsample (see Moretti et al. \cite{moretti2004}). Our BFS are therefore
likely not to be fake detections.  

\subsection{Visual inspection: when a blank is not a blank}

We inspected Digitized Sky Survey 2 (DSS2; McLean et al. \cite{mclean2000}),
Super Cosmos Sky (SSS; Hambly et al. \cite{hambly2001}), 2MASS, FIRST, Sydney
University Molonglo Sky Survey (SUMSS; Bock et al. \cite{bock1999}), NRAO VLA
Sky Survey (NVSS; Condon et al. \cite{condon1998}) and Westerbork Norther Sky
Survey (WENSS; Rengelink et al. \cite{rengelink1997}) images for each of the
201 sources left. Furthermore, we cross-checked source
positions with Simbad, NED and with the VIZIER catalogs (in particular the
USNO-B1 catalog, Monet et al. \cite{monet2003}). 
We excluded from the sample sources and fields already known and well studied
or too complex (e.g. Magellanic Clouds, M31, Orion Nebula). We excluded also
sources positionally coincident with bright stars (holes in the GSC2
coverage), or at the periphery of optically extended emission (more than
$10''$ from the centroid and so escaped the automatic cross-identification). We
found, and excluded, some cases in which our X-ray sources coincided with
reliable USNOB1.0 entries (the criteria for the GSC2 detection are somewhat
more conservative), or with counterparts at other wavelengths (IRAS Faint
Sources Catalog, Moshir et al. \cite{moshir1990}). We prefer not to be too
severe at this stage, deferring strict cuts after the boresight
correction. 72 sources survived this phase.  

\subsection{Cheshire cat: elimination of fluctuations}

We put aside from the primary sample sources which have not been detected
in a longer HRI or ROSAT PSPC pointing (the PSPC had the same passband 
but a substantially larger effective area than the HRI). These could be
variable sources detected at peak flux, but they are also candidates for false
detections. No source has been inserted in this subclass on a non-detection
basis with other instruments, because of the different passband, i.e., given
the absence of spectral information on our sources, the non-detection could be
ought to a peculiar spectral shape. For some sources the suspicion of a false
detection is strengthened by the number of non-detections or by the
respective length of the pointings.  
However, there are cases for which source parameters like detection
probability and number of counts seem to indicate a real source. Furthermore,
as discussed in Section 3.4, we expected only $\sim 2$ spurious detection in
201 sources. Of the 72 sources, 16 were placed in this transient candidate
sub-sample, while for the other 56 there was no evidence of a transient
nature.

\subsection{Pin-pointing the sources to the sky: boresight correction}

We performed some astrometry on the remaining fields (including those of
transient candidates), in order to get rid of the $10''$ fiducial boresight 
uncertainty and therefore to fully exploit the HRI angular resolution capability.
Furthermore, the error given by the detection algorithm (see Section 2.1), 
while not being entirely statistical, can be as a first approximation
treated as random, so that the usual Gaussian relations can be used.
In contrast, the offset given by the boresight uncertainty varies in a random way
in the ensemble of all the pointings, but is systematic in nature for all
the sources in the single pointing, weakening the rejection of distant 
optical associations to BFS.

For each pointing, we matched X-ray sources to optical counterparts (optical 
positions of known X-ray emitters), if any, or to optical catalog sources
distant less than $10''$. 
We excluded X-ray extended sources and sources with no optical catalog entry
in $10''$. Ambiguous cases, i.e. with two or more optical sources present,
were treated individually using the distance and the optical luminosity as
criteria for the identification and then checked a posteriori.
Even if only another X-ray source was present, the shift 
for this source has been applied to the pointing, although obviously in these cases the 
corrected positions and uncertainties have to be taken \textit{cum grano salis}.
Generally, the new error is less than $10''$; 
even if it remains around $10''$, the new position should be free from
systematic boresight uncertainty. Note that performing the boresight
correction only after the selection of Section 3.1 can bring to the loss of
BFS. However, this allows us to deal with a limited number of sources and
boresight corrections.

\section{The data}

The elimination of sources with an off-band counterpart in a $4\sigma$ radius
from the boresight corrected position left us with our final sample of four
sources (three of them transient candidates). 
This $4\sigma$ limit assures us that starting with 1340 sources
(i.e. sources outside the inner $3'$ in the BMW-HRI catalog and obeying to the
cuts above unless th lack of counterparts), only 0.08 of them would lie 
outside the error region by chance. 

In summary, our final BMW-HRI Blank Fields Sources are:
\begin{itemize}
\item Blank: no counterpart at other wavelengths in a $4\sigma$ radius from
the boresight corrected position. 
\item Bright: $\rm f_{X}\geq 2.7\times 10^{-13}$ erg s$^{-1}$ cm$^{-2}$. Note
that from the GSC2 magnitude limit, this is equivalent to $\rm f_{X}/f_{BJ}$
respectively $\geq$ 37 and $\geq$ 59 for Northern and Southern sources. 
\item Point-like: extension significance is 0.
\item Well detected: total (uncorrected) counts $\geq$ 25.
\end{itemize}

After this selection we end up with one persistent BFS plus three transient BFS.
We report the new positions and error radii (as obtained from the boresight
correction procedure), the flux, the detection probability (i.e. the
probability to detect a background fluctuation as a source), the total number
of counts, the Galactic coordinates, the integrated Galactic column density,
the distance from the nearest off-band association (in terms of error radii),
the lower limit on the $\rm f_{X}/f_{BJ}$ and the upper limit on the radio
emission (see Table \ref{finalbmw}).
Note that, though with different depth, all but one (1BMW200739.8-484819) of
the sources have a radio flux upper limit. 

\begin{table*}[htb]
\caption{{BFS parameters.}\label{finalbmw}}
\vspace{0.1 cm}
\footnotesize
\begin{tabular}{|l|c|c|c|c|c|c|c|c|c|c|}
\hline
Name & R.A.  & Dec   & bII & Err. rad.&Flux
&Prob.&Cts&N$_H$& $\rm f_{x}/f_{BJ}$ & Close ass. \\
\hline
 & J2000 & J2000 &&$''$ &erg cm$^{-2}$ s$^{-1}$&$\sigma$&&$10^{20}$ cm$^{-2}$ &
&(sigma)\\
\hline
1BMW042142.4-571541 &04 21 42.6&-57 15 39&-42.54&2.9 &6.5$\times 10^{-13}$&
14.0& 742 & 1.8 & $>141$ & 5.3 \\   
\hline
\hline
1BMW135703.0+181122 &13 57 02.9&+18 11 21& 72.45&7.1 &3.5$\times 10^{-13}$&
 4.2& 112 & 2.1 & $>47$  & 6.3 \\
\hline
1BMW200739.8-484819 &20 07 39.8&-48 48 19&-32.31&5.8 &3.0$\times 10^{-13}$&
 4.3&  55 & 5.1 & $>65$  & 5.2\\ 
\hline
1BMW043306.8+155307 &04 33 06.8&+15 53 07&-21.16&6.5 &2.9$\times 10^{-13}$&
 4.2&  34 & 17.0& $>40$  & 4.9\\ 
\hline
\end{tabular}
\vspace{0.2 cm}
\par\noindent
\end{table*}

\begin{figure}[htbp]
\psfig{figure=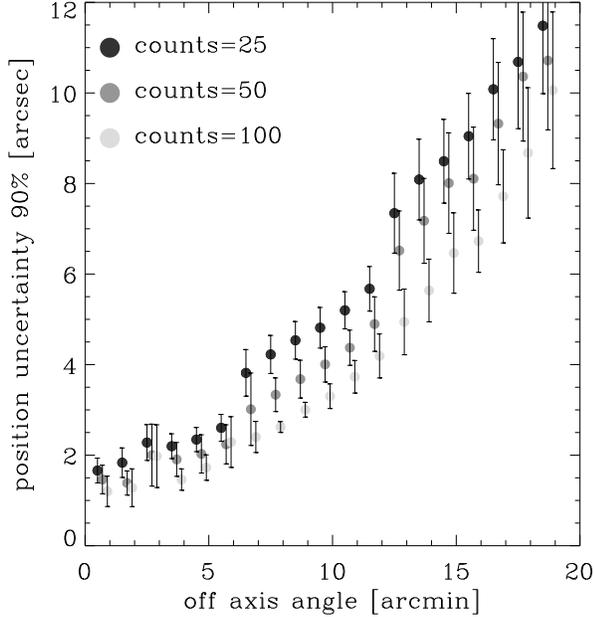,height=9cm}
\caption{BMW-HRI position error: the statistical errors on position of BMW-HRI
sources for different number of counts as a function of the off-axis
angle. Error bars are at $2\sigma$. The change in the slope of the curves
corresponds to angular resolution loss at the change of image rebinning. An
offset of $0.2'$ and $0.4'$ has been applied to points in the 50 counts bin
and 100 counts bin, respectively, for visualization purposes.}
\label{offaxis}
\end{figure}

\subsection{Properties of BFS}

The final error radius estimate for our sources is quite different from case
to case, ranging from $\sim 3''$ to $\sim 7''$; the latter large error is due
to the large off-axis angle and to the low number of counts (see Fig. \ref{offaxis}). In
fact, all our sources are quite offset from the center of the pointing
($>15'$). This depends from the used selection procedure and needs some
comments. First, we concentrate on the consequences of the requirement of
absence of offband counterparts in a $10''$ radius. This affects strongly the
overall distribution in detector coordinates. In fact, the fiducial $10''$
radius for the initial correlation, chosen to match the worst case boresight
error, is huge in comparison to the statistical positional error for the
innermost sources, but it corresponds to the intrinsic error for the outermost
sources (see Fig. \ref{offaxis}). Sources in the innermost part are preferentially
associated with optical counterparts which may not be the true ones, since
they have been searched for over a region much larger than their error
boxes. The overall effect is to deplete the distribution from inner sources,
enhancing the proportion of outer sources. This bias is present independently
of the real size of the boresight correction error. Second, there is a trend
in overestimating outer source counts (especially at low values), since it is
favorable to detect them on top of a positive background fluctuation (an
effect known as Eddington bias, see Hasinger et al. \cite{hasinger1993};
Moretti et al. \cite{moretti2002}). This again adds on the enhancement of
outer sources. The first bias leads to the loss of BFS, but it does not affect
our selected sources. The effect of the second bias on the estimated flux can
be evaluated a posteriori from the duration of the pointings and the number of
counts. Very conservative assumptions give a factor of $\sim 2$ as the maximum
overestimate. 

\begin{figure}[htbp]
\psfig{figure=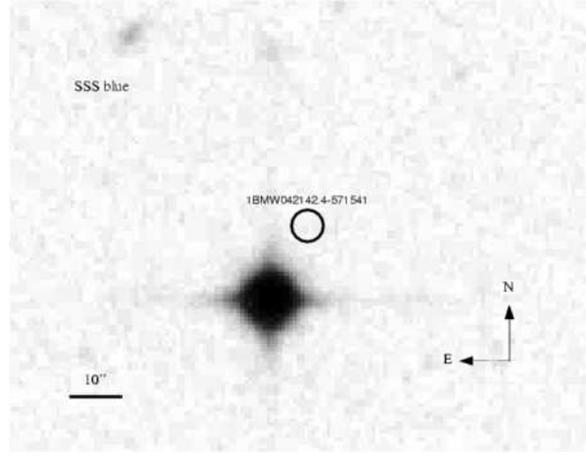,height=6cm} 
\caption{The X-ray position of 1BMW042142.4-571541 superimposed to the
corresponding blue image along with it $1\sigma$ error circles.}
\label{finding1}
\end{figure}

\subsection{1BMW042142.4-571541} 

This source is the brightest of our sample and the one with the
highest number of detected photons. The good statistics involves a
small detection algorithm error radius, so the position
determination is quite accurate (see Fig. \ref{finding1}). The boresight correction was
performed with the known position of the quasar HE 0419-5657 and the nucleus
of NGC 1547 as well as with other eleven sources matched to optical catalogs. 
The source is close ($\sim 15''$ but actually more than 5 $\sigma$) to a
bright star (BJ=13.31, F=11.61 in the GSC2.2); this makes less reliable the
lower limit on the $\rm f_{X}/f_{opt}$, in particular in the red band. 
There are two fake USNO B1.0 sources along the saturation spikes of the bright
star (see Fig. \ref{finding1}). The source was observed with Einstein (1E 0420.7-5723) and
with the ROSAT PSPC (1WGAJ0421.7-5716) and was also detected as a RASS Bright
Source (1RXS J042144.0-571601). The WGA computed flux is $8.8\times 10^{-13}$
erg cm$^{-2}$ s$^{-1}$, consistent with the RASS count rate.
Taking into account the caveats on the count rate to flux conversion, this
could mean that there is no evidence for long term flux variations.

\begin{figure}[htbp]
\psfig{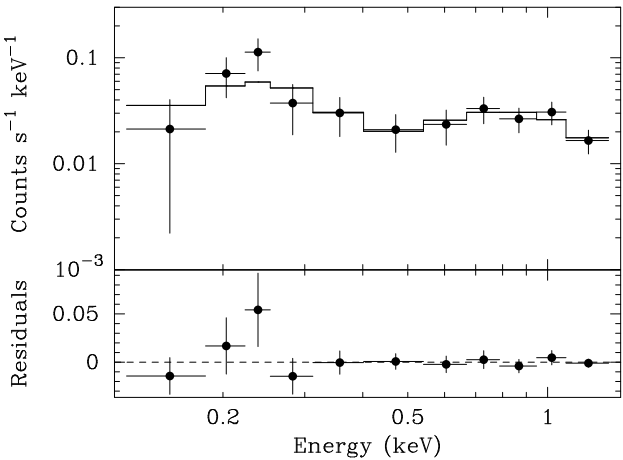}
\caption{Power law spectral fit of the PSPC observation of 1BMW0421.4-571541.}
\label{figpspc}
\end{figure}

We re-extracted the archival PSPC observation of this source (sequence
rp700034n00). The source was very close to the PSPC rib; despite this, we
attempted a rough spectral analysis. We extracted source photons from a circle
of radius $200''$ centered on source position. We used as background region a
circle of radius $240''$, centered in an empty region south of the source. The
collected counts were 321. We rebinned channels by a variable factor in order
to have at least 20 photons in each bin. Channels 1-11 and 136-256 were
ignored. Figure 3 shows the resulting spectrum, which we then fitted with
XSPEC (v.11.2). Due to the poor statistics, we considered a limited number of
models, as shown in Table \ref{pspc}. The spectrum is definitely soft, without
evidence of absorption excess. The best fit absorbed blackbody model tends to
null absorption, with a $90\%$ upper limit on column density of $\sim 3\times
10^{19}$ atoms cm$^{-2}$. The power law fit gives a relatively better reduced
$\chi^2$ (1.3 versus 1.4). Furthermore, the best column density value is quite
close to the integrated Galactic value of $1.2\times 10^{20}$ atoms cm$^{-2}$
(see Fig. \ref{figpspc}).
Freezing absorption at the Galactic value, the reduced $\chi^2$ does not
change appreciably. The bremsstrahlung fit gives the best reduced $\chi^2$
(1.3) and the column density is $\sim 1.3\times 10^{20}$ atoms cm$^{-2}$.

\begin{table}[htb]
\caption{{Spectral fits for PSPC observation of 1BMW042142.4-571541.}\label{pspc}}
\vspace{0.1 cm}
\footnotesize
\begin{tabular}{|l|c|c|c|c|c|}
\hline
Model    & N$_H$           & kT /$\Gamma$     & Flux(0.1-2 keV) & $\chi^2_{\rm
red}$\\
\hline
         &$10^{20}$ cm$^{-2}$& keV             &$10^{-13}$ cgs     & \\
\hline
B.body   &$0^{+0.3}$        &$0.19^{+0.02}_{-0.02}$& 5.6      & 1.4\\
\hline
B.body   &1.8(frozen)     &$0.15^{+0.02}_{-0.01}$& 6.9      & 1.9\\
\hline
Pow.-law &$1.8^{+1.8}_{-1.4}$ &$1.8^{+0.7}_{-0.6}$& 6.5     & 1.3\\
\hline
Pow.-law &1.8(frozen)     &$1.8^{+0.2}_{-0.2}$& 6.5     & 1.2\\
\hline
Bremss.  & $1.3^{+1.0}_{-0.9}$&$1.46^{+14}_{-0.77}$& 6.2     & 1.3\\
\hline
Bremms.  &1.8(frozen)     &$1.03^{+0.49}_{-0.25}$& 6.2     & 1.2\\
\hline
\end{tabular}
\vspace{0.2 cm}
\par\noindent
\end{table}

\subsection{Transient BFS}
In this section are presented three sources (see Table \ref{finalbmw}) that,
while fulfilling all the others criteria for BFS, show additional strong
evidence of a transient behaviour.    
In fact, despite their brightness in the BMW-HRI catalog
($\rm f_{X}\geq 2.7\times 10^{-13}$ erg s$^{-1}$ cm$^{-2}$), for
each of them there is at least another ROSAT HRI or PSPC
pointing where they have not been detected (deeper than the
HRI exposure that revealed them). 

\begin{figure}[htbp]
\psfig{figure=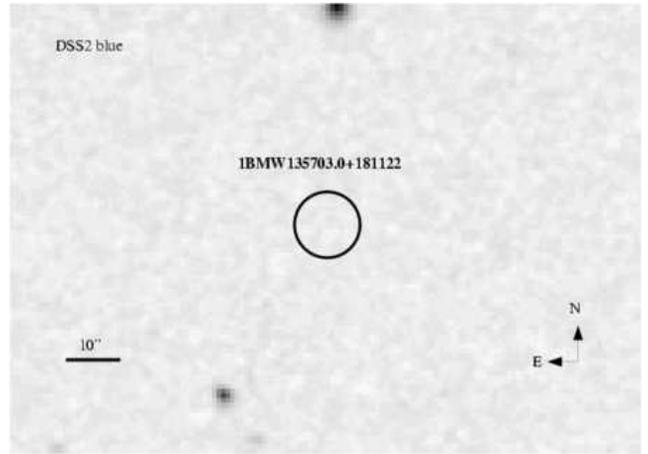,height=6cm} 
\caption{The X-ray position of 1BMW042142.4-571541 superimposed to the
corresponding blue image along with it $1\sigma$ error circles.}
\label{finding2}
\end{figure}

\textbf{1BMW135703.0+181121.}
1BMW135703.0+181121 has been detected in a $\sim$ 5.5 ks HRI pointing in Jan.
1998; we included it in the possible transient sub-sample since it has not
been detected in three longer PSPC pointings ($\sim $11.4 ks, Jan. 1992;
$\sim$8 ks, Jul. 1990; $\sim$5.6 ks, Jul. 1992; also it was not seen in a
shorter $\sim$1.1 ks HRI pointing in Jan. 1992). The high number of counts
detected points toward a real, transient source. The boresight corrected
position has been obtained with three sources. There are no counterpart at
other wavelength in a 45$''$ radius ($\sim 6.3\sigma$, see Fig. \ref{finding2}).

\begin{figure}[htbp]
\psfig{figure=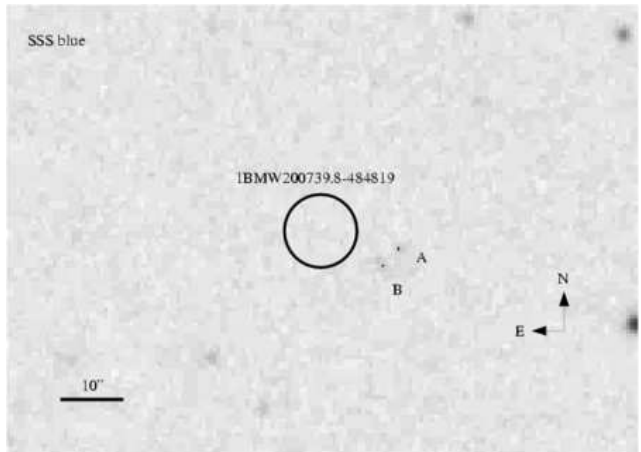,height=6cm} 
\caption{The X-ray position 1BMW200739.8-484819 superimposed to the
corresponding blue image along with it $1\sigma$ error circles.}
\label{finding3}
\end{figure}

\textbf{1BMW200739.8-484819.}
This source has been detected in a $\sim$ 3.2 ks pointing in Oct. 1996, while
no corresponding source was observed in four other pointings of
comparable length (2.6-3.6 ks) at distance of $\sim$ days, nor in two $\sim$11
ks ROSAT PSPC pointings in Nov. and Apr. 1992. The closest optical catalog
source is at more than 30$''$ (see Fig. \ref{finding3}) from the boresight corrected X-ray
position (the boresight correction has been performed with the other two X-ray
sources detected in the Oct. 1996 HRI pointing). Two objects, A and B in
Fig. \ref{finding3}, are clearly visible in the blue SSS image, respectively at $\sim16''$
($\sim 2.7\sigma$) and $\sim 13''$ ($\sim 2.3\sigma$). Object B (BJ=22) is
reported to be extended, object A (BJ=22.3) is possibly extended too. Oddly,
while nothing is apparent at that position in the red SSS image, a source can
be seen both in the shallower ESOI RED image and in the SSS I image, at a
position almost coincident with object B, with RF=20.2 (more than 2 mag
brighter than in the Schmidt plate limit) and IN=19.0. An obvious hypothesis
could be the presence of an optically variable source, that perhaps could be
identified with 1BMW200739.8-484819.

\begin{figure}[htbp]
\psfig{figure=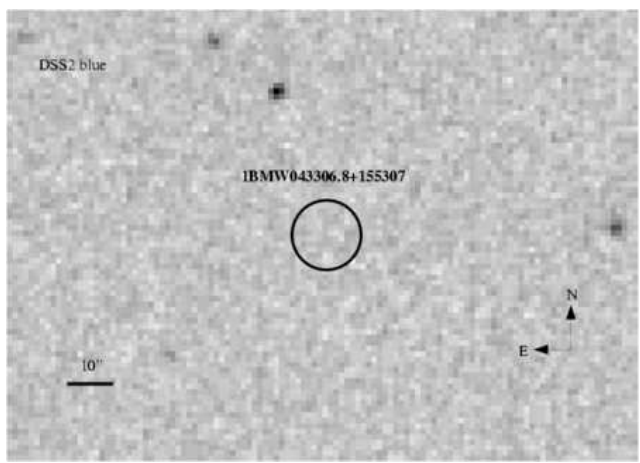,height=6cm}
\caption{The X-ray position 1BMW043306.8+155307 superimposed to the
corresponding blue image along with it $1\sigma$ error circles.}
\label{finding4}
\end{figure}

\textbf{1BMW043306.8+155307.} The observation in which this source was
detected ($\sim$2 ks in Feb. 1998) is part of the Hyades ROSAT campaign. The
source is included among transients due to 
its non-detection in four PSPC pointings (of $\sim$2.4 ks, Feb. 1991;
$\sim$1.9 ks, Mar. 1991; $\sim$16.6 ks, Sep. 1992; $\sim$19.6 ks, Aug. 1993).
In all the four PSPC pointings the source was at much larger off-axis angle
than in the HRI one ($31.5'-55.1'$ versus $15.8'$). The boresight correction
was done with three sources. There are no possible counterparts in a $\sim
25''$ ($\sim 3.8\sigma$) radius from the source position, where there is a
source detected only in the APM North Catalog (E=19.8), not visible in the
DSS2 digitized sky images. The other closest source is at $32''$ ($\sim
4.9\sigma$, see Fig. \ref{finding4}).

\section{Discussion: nature of the blank fields}


In order to proceed further with hypotheses on BFS nature we need to compare
the $\rm f_{X}/f_{BJ}$ limits of BFS with existing classification schemes. We
chose as a reference scheme the one of the RASS-Hamburg optical identification
program (Zickgraf et al. \cite{zick2003}), which uses similar X-ray (0.1-2.4
keV) and optical (Johnson B) bands. 
We converted the RASS-Hamburg limits (see Table 6 of Zickgraf et
al. \cite{zick2003}) to BMW-HRI and GSC2 $\rm f_X/f_{BJ}$ (the deepest GSC2
band), keeping into account the different X-ray energy band and spectral shape
used for flux evaluation, and the different optical band. The resulting
classification scheme is shown in Table \ref{fxfbj} (see Chieregato
\cite{chie2005} for details about the conversion).  
The $\rm f_{X}/f_{BJ}$ lower limits of BFS, $\simeq$ 37 in the
Northern sky and $\simeq$ 59 and in the Southern sky (due to different GSC2
depth, see Table \ref{crossid}), are a factor of $\sim 2$ and $\sim$ 3 beyond
the limits of our reference values for different classes, respectively. 
This points out the possibility that BFS are unusual objects,
left outside from classifications, or at least very peculiar members 
of the more ordinary categories. 

\begin{table}[htb]
\caption{{$f_{X}/f_{BJ}$ for various classes of sources, adapted to BMW-HRI
fluxes and GSC2 BJ band.}\label{fxfbj}} 
\vspace{0.1 cm}
\begin{tabular}{|l|c|c|c|r|}
\hline
Class        & Lower limit     & Upper limit     & B-V\\
\hline
 Stars       &1.3$\times 10^{-5}$ &0.6             &-0.5$\div$2\\
 \hline
 White Dwarfs&1.2$\times 10^{-5}$ &9.3$\times 10^{-4}$&-0.3$\div$1.5 \\
 \hline
 Cat. var.   &1.8$\times 10^{-2}$ &4.6             &-0.1$\div$1 \\
\hline
Galaxies     &3.1$\times 10^{-2}$ &11.1            & 0.4$\div$1.5 \\
\hline
Gal. clusters&0.2              &16.9            & 1.0$\div$1.5\\
\hline
AGN (w. BLLacs)&0.22          &12.0            &-0.5$\div$1 \\
 \hline
\end{tabular}
\vspace{0.2 cm}
\par\noindent
\end{table}

\subsection{BFS as unusual sources}

It is well known that some rare class of sources can reach extreme values of
${\rm f_X/f_{opt}}$ (see Cagnoni et al. \cite{ila2002}). Here we will focus on
our specific sample and discuss the possibility that
our BFS belong to the following categories:
\begin{enumerate}
\renewcommand{\theenumi}{\alph{enumi}}
\item Isolated Neutron Stars (INS).
\item X-ray binaries.
\item High Redshift or Dark Clusters of Galaxies.
\item Type 2 Quasars.
\item Extreme BL Lac.
\end{enumerate}

\textbf{Isolated Neutron Stars.}
Isolated Neutron Stars (INS; see Treves et al. \cite{treves2000}, Haberl
\cite{haberl2003}, Haberl \cite{haberl2004}) are extreme BFS. The bona-fide
$\rm f_{X}/f_{opt}$ lower limit used to assess their nature is 1000, but for
optically identified objects the real value can be as high as $10^5$. The
X-ray spectrum is optimally fitted by a black body with kT$\sim$60-100 eV.
The optical emission lies a factor of a few over the Rayleigh-Jeans
tail of the blackbody and no emission outside the optical/soft X-ray range has
been found. The low column densities from the X-ray spectra and the parallax
measured distance for RXJ1856.5-3754 (Walter \& Lattimer \cite{walter2002})
makes this source an intrinsecally faint and closeby object, probably emitting
from the neutron star surface.\\  
Different kind of isolated neutron stars can also reach high
$\rm f_{X}/f_{opt}$ values: Anomalous X-ray Pulsars (AXP; Israel, Mereghetti
\& Stella \cite{isra2002}), close-by radio pulsars with substantial cooling
X-ray thermal emission (e.g. PSR 0656+14) and Geminga-like objects (Bignami \&
Caraveo \cite{bigna1996}).  AXPs are young, luminous and distant objects and
should be much closer to the Galactic plane than our BFS. The radio limit,
though not strong, argue against close-by cooling radio pulsars. This does not
apply to Geminga-like neutron stars for which the radio emission is only
marginal. Converting the X-ray flux of Geminga to the HRI pass-band and using
V=25.6, yields ${\rm f_X/f_V}\sim 67$, making it concistent with our BFS candidates. In
the cooling framework, the same physical mechanism is at the basis of INS
emission and of the soft thermal component of young radio pulsars and
Geminga-like objects. 
Basing on this scenario, Popov et al. (\cite{popov2003}, \cite{popov2004})  
produced the log N-log S of cooling neutron stars (all the three categories),
allowing us to estimate that only $\sim$ 1 unidentified INS is expected for
our flux and sky coverage. 
If our steady BFS (1BMW042142.4-571541)) is an INS, no optical counterpart
could be revealed up to the $\sim$ 28 mag; therefore, this possibility
could be strengthened by the absence of optical sources in deeper
observations, reaching the critical $\rm f_{X}/f_{opt}$ lower limit of 1000,
corresponding to magnitude $\sim$ 25-26, together with a better position from
Chandra.    
At the same time the well-defined properties of the X-ray emission
could be tested by observations with an instrument with sufficient soft
response, like the XMM-Newton EPIC pn camera. 

\textbf{X-ray binaries.}
Some kind of X-ray binaries can reach high values of $\rm f_{X}/f_{opt}$.
In particular, BFS searches have already discovered Ultraluminous X-ray Sources
(ULX; e.g. Cagnoni et al. \cite{ila2003}). However, the observed flux level of
BFS is too low for a Galactic object and none of our BFS can
belong to nearby galaxies, excluding ULXs, supersoft sources and bursting
LMXBs. Therefore, we consider unlikely known classes of X-ray binaries (see
however the discussion on possible transients).

\textbf{Clusters of Galaxies.}
BFS research has already been effective for the discovery of clusters of
galaxies since Cagnoni et al. (\cite{ila2001}). Furthermore, several distant 
clusters have already been discovered in the BMW-HRI catalog as extended
objects (Moretti et al. \cite{moretti2004}).  
Distant clusters may escape this selection if only the X-ray peak is bright
enough to be detected, resulting in a point source, while the rest is
concealed in the background.  
This effect is strongly favoured at large off-axis angle (and consequent PSF
degradation) like those at which BFS are detected. Therefore, if some BFSs
are galaxy clusters, we expect them to be at quite high redshift. The $\rm
f_X/f_{opt}$ of clusters of galaxies is the highest among  
the X-ray emitting classes of Table \ref{fxfbj}. For normal galaxy clusters, the values
reached are usually lower than those of BFS (and if only the peak of the X-ray
emission is computed, the resulting $\rm f_X/f_{opt}$ is further lowered).
Despite this, there are two effects that could boost the $\rm
f_X/f_{opt}$. The first, and more effective, 
is the redshift of the optical emission. In fact, for a redshift of 0.75, the
4000 \AA\ break of the cD galaxy (or of any early type galaxy of the cluster)
would be redder than the GSC2 F band. Even a lower redshift could result in a
sufficient enhancement of the $\rm f_X/f_{opt}$. 
Second, the X-ray peak could be not coincident with the optical one (if the
cluster is not virialized, or if there is no cD galaxy, like in Bautz-Morgan
type III clusters, or if there is a central cooling flow). 
Optical-near IR deeper observations should reveal a substantial increase of
galaxy counts in proximity of BFS and, if there is any, the cD galaxy of the
cluster. An even more strong test would be X-ray observations with Chandra or
XMM-Newton, that should reveal an extended source.

\textbf{BL Lacs.}
 BL Lacs stand somehow apart from the bulk of AGNs in that
they can reach high ${\rm f_X/f_{opt}}$ values.
The unified spectral energy distribution of BL Lacs (Fossati et
al. \cite{fossati1998}; Ghisellini et 
al. \cite{ghis1998}) can be described by a two peak emission, in which the
lower energy peak (radio to X-ray) is commonly ascribed to synchrotron
emission, while the high energy peak (X-ray to $\gamma$-ray) is probably due to
inverse Compton emission. In this scenario, different BL Lacs are
distinguishables on the basis of the shifts of the peaks, creating the
so-called blazar sequence. By tuning the position of the peaks, it is possible to
produce the old categories of X-ray selected and radio selected BL Lacs, the
former being those capable of high ${\rm f_X/f_{opt}}$ values. 

In order to check the viability of BL Lacs hypothesis for BFS we consider 
the different $\alpha_{xo}$ (the power law index connecting the X-ray flux to
the optical flux) and $\alpha_{xr}$ (the power law index connecting the X-ray
flux to the radio flux) for Einstein Slew Survey Bl Lacs (Perlman et
al. \cite{perlman1996}) and the upper limit for BFS.   
Despite the poor radio limits, the locus occupied by BFS is partially
superimposed to the most extreme Slew BL Lacs, with the brightest BFS,
1BMW042142.4-571541, exhibiting an $\alpha_{xo}$ limit substantially lower.   
If these objects were indeed BL Lacs, they would probably stretch the sequence
in the $\alpha_{xo}$-$\alpha_{xr}$ plane, representing a still undiscovered
extreme population. 
The identification of the optical and possibly of the radio
counterpart would be very important for testing the BL Lac hypothesis.
In this case, further X-ray observations should show a power law spectrum with
photon, without evidence of absorption excess.

\textbf{Type 2 Quasars.}
Evidences for a substantial population of type 2 Quasars, as requested by
unification models (Urry \& Padovani \cite{urrypd1995}; Comastri et
al. \cite{comastri1995}), has been continuously increasing in the recent
past. In particular, the resolution of the largest fraction of the
cosmic X-ray background in discrete sources (Hasinger et al. \cite{has1998};
Mushotzky et al. \cite{mush2000}; Campana  et al. \cite{camp2001},
Giacconi et al. \cite{giacconi2002}, Moretti et al. \cite{moretti2003}, Brandt
\& Hasinger \cite{bra2005}) has brought to infer the existence of an
adequate number of highly absorbed and luminous AGNs, responsible for the hard
part of the background. 
Up to now, while only a few type 2 quasars have been firmly identified
(e.g. Stern et al. \cite{stern2002}; Norman et al. \cite{norm2002}),
many candidates are being produced either by deep fields, pencil beam searches
(Chandra Deep Field North, Barger et al. \cite{barger2003}; Chandra Deep Field
South, Szokoly et al. \cite{szokoly2003}; Lockman Hole, Mainieri et
al. \cite{mainieri2002}), or by dedicated shallower surveys, with a
substantial sky coverage (Hellas2XMM, Baldi et al. \cite{baldi2002}; ChaMP,
Green et al.\cite{green2003}; SEXSI, Harrison et al. \cite{harrison2003};
BMW-Chandra, Romano et al. \cite{romano2004}). 
The most striking characteristic of type 2 quasars is the severe
absorption excess. In particular, this makes them Extremely Red Objects, with
R-K$>5$. 
Therefore, if some BFS are type 2 quasar, we expect them to have a relatively bright
near-IR counterpart. Unfortunately, the near-IR limit of the 2MASS (Table \ref{crossid})
is not sufficient to confirm or rule out this possibility. Deeper
near-IR observations are therefore mandatory to test the viability of this
scenario. Future X-ray spectral information could also be extremely useful,
confirming or excluding the need of absorption excess to fit the spectra.
A number of source capable of very high ${\rm f_X/f_{opt}}$ ratios was
recently discovered in Chandra Deep Fields, the so-called Extreme X-ray
Objects (EXO; Koekemoer et al. \cite{koe2004}). The nature of these sources is
still not clear, but they share with type 2 quasars the extremely red
colour. However, these sources have X-ray fluxes much fainter than BFS ($\le
4\times 10^{-15}$ erg cm$^{-2}$ s$^{-1}$), so they are likely to belong to
different populations.

\bigskip

The nature of BFS reported here is probably not-unique. 
In order to accomodate the possible transients in one of the above
classes, a key factor is variability. The most variable candidates,
BL Lacs, can vary up to a factor of $\sim 100$ (e.g. Mkn 501; Pian et
al. \cite{pian1998}), enough to explain the non-detections. This makes BL
Lacs appealing candidates for possible transients. The variability of type 2
quasars, instead, seems not sufficient. Other known transient source classes
usually can not produce such high $\rm f_{X}/f_{opt}$ values.
An alternative explanation could be
that of a new kind of transient, capable of great flux variability but reaching
much lower peak maximum luminosity. According to this idea, 
the $\rm f_{X}/f_{opt}$ would be greatly boosted by the non-simultaneity of
X-ray and optical observations.  

\section{Follow-up projects}

We wish to stress that the interpretation of BFS requires future
multi-wavelength observations. Deeper optical and near-IR observations should
allow us to find the optical counterpart (except possibly for INS), and future
X-ray observations should give spectral and temporal insights. We already
started a program of optical follow-up. In the future, more will be known 
on these mysterious and fascinating lacunae in the X-ray/optical sky.

\acknowledgement
{We thank Ilaria Cagnoni for inspiration and valuable help; Marina Orio and
Nicola La Palombara for useful discussions. We thank the referee, Gianni
Zamorani, for his constructive criticisms which improved considerably the
manuscript and his patience in the review process. We acknowledge usage of
HEASARC and CDS databases and tools. This work was partially supported by
MIUR under contract COFIN 2002027145\_002.
}
\endacknowledgement

\end{document}